\documentclass[aps,preprintnumbers,amsmath,amssymb,usenatbib,nofootinbib,superscriptaddress,showkeys,showpacs,10pt]{revtex4-2}
\usepackage{amsmath}
\usepackage{amsfonts,color}
\usepackage{amssymb,float}
\usepackage{graphicx}
\usepackage{subfigure}
\usepackage{appendix}
\usepackage[colorlinks]{hyperref}
\usepackage{xcolor}
\usepackage{orcidlink}
\usepackage{epsf}
\usepackage{bm}
\usepackage{epstopdf}
\usepackage{natbib}
\usepackage{lipsum}
\usepackage{multirow}
\usepackage{nicematrix}
\usepackage{booktabs}
\usepackage{array}

\begin{document}

\title{Interacting Holographic Dark Energy in $f(Q)$ Gravity: Cosmological Evolution and Gravitational Wave Signatures}

\author{N. Amiriborkhani}
\email{n.amiriborkhani@iau.ac.ir}
\affiliation{Department of Physics, Am.C., Islamic Azad University, Amol, Iran.}

\author{Alireza Amani\orcidlink{0000-0002-1296-614X}}
\email{al.amani@iau.ac.ir}
\affiliation{Department of Physics, Am.C., Islamic Azad University, Amol, Iran.}

\author{M. A. Ramzanpour}
\email{mohammad1352@iau.ac.ir}
\affiliation{Department of Physics, Am.C., Islamic Azad University, Amol, Iran.}

\date{\today}
\begin{abstract}

In this paper, an interactive Holographic Dark Energy (HDE) model is studied in the framework of modified gravity \(f(Q)\). By adopting a power parameterization for the Hubble parameter, the field equations are reconstructed and the evolution of the universe at the background level and tensor perturbations are investigated. Then, using observational data \(H(z)\), the model parameters are constrained and the dynamical behavior of dark energy throughout the history of the universe is analyzed. Also, the study of the evolution of energy density, pressure and the Equation of State (EoS) parameter of dark energy shows that dark energy in the late universe naturally tends to a region close to the cosmological constant behavior, while in the past it followed a distinct dynamical evolution. Stability analysis based on the speed of sound also indicates that the model has good classical stability around the present era. In addition, the compatibility of the current values of the relative density parameters of matter and dark energy with the observational constraints confirms the ability of the model to reproduce the main features of the observed universe. Next, the propagation of gravitational waves in the cosmological context of the model is investigated. The results show that the corrections due to \(f(Q)\) gravity and the interaction between matter and dark energy can affect the evolution of tensor perturbations and produce signatures distinct from the standard scenario. Overall, the findings of this study indicate that the interactive HDE in the gravitational framework \(f(Q)\) can provide a consistent framework for describing the cosmic acceleration and studying the cosmological consequences of gravitational waves.

\pacs{98.80.-k, 98.80.Es, 95.35.+d, 04.30.-w}

\keywords{Gravitational waves; Interacting $f(Q)$ gravity; Holographic dark energy; Dark energy; Dark matter}

\end{abstract}

\maketitle
\newpage
\tableofcontents

\section{Introduction}\label{sec1}

The discovery of the accelerating expansion of the Universe through observations of Type Ia supernovae \cite{Riess1998, Perlmutter1999} is considered a landmark in modern cosmology. This result was later confirmed by a series of independent observations, including the Cosmic Microwave Background (CMB) \cite{Bennett2003, Caldwell2004}, Baryonic Acoustic Oscillations (BAO) \cite{Eisenstein2005, Percival2010}, galaxy surveys such as the Sloan Digital Sky Survey (SDSS) \cite{Riess2004, oka2014simultaneous}, data from the Wilkinson Microwave Anisotropy Probe (WMAP) mission \cite{Spergel2003}, and studies of the Large-Scale Structure (LSS) of the Universe \cite{Koivisto2006, Daniel2008}. Within the framework of the standard \(\Lambda\)CDM model, the current acceleration of the universe is attributed to the presence of a negatively charged energy component called dark energy, which is described in the simplest case by the cosmological constant \(\Lambda\) \cite{weinberg1989cosmological}. Despite the success of this model in fitting a wide range of observational data, theoretical problems associated with the cosmological constant, including the fine-tuning problem and the coincidence problem, as well as some recent observational tensions, have provided significant motivation to investigate alternative descriptions of the cosmic acceleration. Meanwhile, modified gravity theories have been proposed as one of the most important approaches beyond general relativity, in which the origin of the acceleration of the universe can come from modifying the geometric part of gravitational action \cite{chiba2000kinetically, kamenshchik2001alternative, bagla2003cosmology, amani2013interacting, dolgov2003can, iorio2016constraining, faraoni2006matter, nojiri2006introduction, jamil2010new, jawad2013reconstruction, capozziello2011cosmography, myrzakulov2011accelerating, mirzaei2020observational, pourbagher2020thermodynamics, harko2011f, nojiri2005modified, khlopov1985gravitational, dymnikova2000decay}.

Among modified gravity theories, symmetric teleparallel gravity has attracted considerable attention in recent years. Unlike general relativity, which attributes gravity to the curvature of space-time, as well as teleparallel gravity, which relies on torsion, this geometric framework is based on nonmetricity, while both curvature and torsion are completely removed \cite{Nester1998, Adak2006, jimenez2018coincident, Lazkoz2019, Xu2019, Frusciante2021, Ayuso2021, Bahamonde2022, Heisenberg2023}. In this theory, the gravitational dynamics is described by the anisometric scalar $Q$, and the linear selection $f(Q)=Q$ leads to equations equivalent to general relativity. The generalization of this structure to an arbitrary function $f(Q)$ gives rise to the theory of $f(Q)$, which today is known as one of the most promising geometric extensions of gravity \cite{jimenez2018coincident, Lazkoz2019}. One of the outstanding features of this theory is that, unlike gravity $f(R)$, its field equations remain second-order and at the same time provide enough freedom to describe the dynamics of the universe on large scales. As a result, gravity $f(Q)$ has been widely studied in various fields including late cosmic acceleration, dark energy, universe expansion history, cosmological stability and gravitational disturbances and has shown promising results.

In addition to modified gravitation, another approach to explaining the nature of dark energy stems from the fundamental ideas of quantum gravity. Among them, the holographic principle, first proposed in studies of black hole thermodynamics and quantum gravity, states that the number of degrees of freedom of a region of space is proportional to the area of its boundary, not to its volume \cite{tHooft1993, Susskind1995}. Based on this principle, Cohen and colleagues showed that the vacuum energy in a quantum field theory must be limited in such a way that the formation of a black hole on the infrared scale is prevented \cite{Cohen1999}. This idea eventually led to the introduction of the HDE model, in which the density of dark energy depends on an infrared length scale \cite{Li2004}. Since this model is rooted in fundamental considerations of quantum gravity, it describes dark energy not as a mere cosmological constant, but as a dynamical component whose evolution is related to the geometry and cosmological horizons. In recent years, various HDE models have been extensively investigated from theoretical and observational perspectives, and the results have shown that this framework is able to describe the expansion history of the universe and the properties of dark energy with good consistency \cite{Li2004, Wang2017, Hsu2004, Huang2004, Wang2017HZ, Nojiri2017, li2004model, hao2009entropy, amani2012logarithmic, del2011holographic, hu2015holographic, Wang2005}.

Considering the independent motivations arising from modified gravities and HDE, the combination of these two frameworks has become an active topic in theoretical cosmology in recent years \cite{amani2015logarithmic}. In this regard, various HDE models have been investigated in the context of $f(Q)$ gravity, and it has been shown that nonmetricity effects can significantly affect the dynamical evolution of dark energy and the expansion history of the universe \cite{Mandal2020, Anagnostopoulos2021, Frusciante2021, Ayuso2021}. In addition, the presence of an interaction between dark matter and dark energy has attracted much attention as an attractive phenomenological possibility, since such an interaction could partially alleviate the cosmic coincidence problem and significantly change the dynamics of the late universe \cite{Amendola2000, Wang2016, Bolotin2015}. From this perspective, studying the HDE of interaction in the $f(Q)$ framework not only allows for the simultaneous investigation of geometric and thermodynamic effects, but also provides a suitable platform for testing the cosmological implications of the theory at different scales.

We can now turn to one of the main motivations of the paper, namely the connection between gravity $f(Q)$, HDE and gravitational waves. In recent years, after the detection of the event GW170817 and its electromagnetic counterpart \cite{Abbott2017, Abbott2017EM}, gravitational waves have become a powerful tool for testing theories of gravity beyond general relativity. Measurements of the very small time difference between the gravitational signal and the electromagnetic radiation showed that the propagation speed of gravitational waves coincides with the speed of light with very high precision, a result that has imposed severe constraints on many modified theories of gravity \cite{Baker2017, Langlois2018, Nojiri2018, maity2024gravitational, debnath2021gravitational}.

In this context, the study of tensor perturbations and the propagation of gravitational waves has become one of the fundamental tests for assessing the validity of dark energy and modified gravity models. Although many theories may be able to reproduce the history of the expansion of the universe at the level of the cosmological background, their behavior in the perturbation part, and in particular in the gravitational wave propagation equation, can lead to significant differences from the Standard Model. Indeed, even in cases where the propagation speed of gravitational waves remains the same as the speed of light, gravitational corrections can leave observable observational signatures through changes in the effective friction, the wave amplitude, the dispersion relation, or the evolution of tensor perturbations \cite{Baker2017, Langlois2018, Nojiri2018, Cai2018}.

On the other hand, in the framework of $f(Q)$ gravity, the tensor perturbation part and the propagation of gravitational waves have not yet been studied as much as the studies on the evolution of the background \cite{Frusciante2021, Atayde2021}. This becomes especially important in the presence of HDE and the interaction between dark matter and dark energy, since both components can affect the expansion rate of the universe and, consequently, the evolution of gravitational waves on cosmological scales. Therefore, simultaneously examining the behavior of the cosmic background and the propagation of gravitational waves in $f(Q)$-based interacting HDE models provides a good opportunity to test the consistency of this framework with current and future observations.

Motivated by the above considerations, in this paper we investigate an interactive HDE model in the gravitational framework \(f(Q)\). To this end, by adopting a power law parameterization for the Hubble parameter, we first solve the cosmological field equations and reconstruct the corresponding gravitational function. Then, using observational data \(H(z)\), we constrain the free parameters of the model and study the evolution of the main cosmological quantities including the deceleration parameter, density and pressure of dark energy, the EoS parameter, the speed of sound, and the relative density parameters of matter and dark energy. Also, the age of the universe and the consistency of the model with current observational values are investigated. Next, tensor perturbations and gravitational wave propagation in the cosmological field obtained from the model are analyzed and the effects of nonmetricity and interaction between dark components on the evolution of gravitational waves are evaluated. The results of this study show that the interactive HDE in gravity \(f(Q)\) can provide a suitable platform for simultaneously describing cosmic acceleration and phenomena related to gravitational waves.

This paper is organized as follows:

In Section \ref{II}, the theoretical framework of gravity \(f(Q)\) is introduced. In Section \ref{III}, we present the HDE model. In Section \ref{IV}, the correspondence between HDE and gravity \(f(Q)\) is established and the corresponding model is reconstructed. Section \ref{V} is devoted to constraining the model parameters using \(H(z)\) data and analyzing its cosmological properties. In Section \ref{VI}, the behavior of tensor perturbations and gravitational wave propagation in this framework is investigated. Finally, Section \ref{VII} is devoted to summarizing the results and presenting conclusions.


\section{Foundation of $f(Q)$ gravity}\label{II}

In this section, we consider a theory based on a torsion-free and non-curvature geometry, called symmetric teleparallel geometry, which is introduced as $f(Q)$ gravity where $Q$ is a non-metricity scalar. In this regime, Ricci curvature and torsion are assumed to be zero, and only non-metricity plays a role. In that case, we write the action of $f(Q)$ gravity as
\begin{equation}\label{act1}
  S = \int d^4x \sqrt{-g} \left( \frac{1}{2 \kappa^2} f(Q) + \mathcal{L}_m \right),
\end{equation}
where $f(Q)$ is an arbitrary function of the non-metricity scalar $Q$, $\mathcal{L}_m$ is the matter Lagrangian density, $g = det(g_{\alpha\beta})$, and $\kappa^2 = 8\pi G $. the non-metricity scalar write down as
\begin{equation}\label{Q1}
  Q = Q_{\alpha \mu \nu} P^{\alpha \mu \nu},
\end{equation}
where the superpotential tensor $P_{\,\,\mu\nu}^{\lambda}$ is non-metricity conjugate which is written by
\begin{eqnarray}
&Q_{\alpha \mu \nu} = \nabla_\alpha \, g_{\mu \nu} = \partial_\alpha \, g_{\mu \nu} - \Gamma^{\lambda}_{~ \alpha \mu} \, g_{\lambda \nu} - \Gamma^{\lambda}_{~ \alpha \nu} \, g_{\mu \lambda}, \label{Palpha1-1}\\
&P^{\alpha \mu \nu} = - \frac{1}{4} Q^{\alpha \mu \nu} + \frac{1}{2} Q^{(\mu \nu) \alpha} + \frac{1}{4} \left( Q^{\alpha} - \tilde{Q}^{\alpha}\right) g^{\mu \nu} - \frac{1}{4} g^{\alpha ( \mu} Q^{\nu)},\label{Palpha1-2}
\end{eqnarray}
where the corresponding traces of the non-metricity tensor are
\begin{equation}\label{Qtilde1}
Q_{\alpha} = g^{\sigma \lambda} Q_{\alpha \sigma \lambda} ~~~~ \tilde{Q}_{\alpha} = g^{\sigma \lambda} Q_{\sigma \alpha \lambda}.
\end{equation}

By varying the action \eqref{act1} with respect to metric tensor, the Einstein equation is obtained as
\begin{equation}\label{Ein1}
\frac{2}{\sqrt{-g}} \nabla_\alpha
\left( \sqrt{-g}\, f_Q\, P^{\alpha}{}_{\mu\nu} \right)
+ \frac{1}{2} g_{\mu\nu} f
+ f_Q \left(
    P_{\mu\alpha\beta}\, Q_{\nu}{}^{\alpha\beta}
    - 2\, Q_{\alpha\beta\mu}\, P^{\alpha\beta}{}_{\nu}
  \right)
= -\,\kappa^2\, T_{\mu\nu},
\end{equation}
where $f_Q = df / dQ$, and $T_{\mu \nu}$ is the energy-momentum tensor.

In what follows, we consider the homogeneous and isotropic Friedmann-Lema\^{i}tre-Robertson-Walker (FLRW) metric
\begin{equation}\label{met1}
ds^2 = -dt^2 + a^2(t) (dx^2+dy^2+dz^2),
\end{equation}
where $a(t)$ is the scale factor. In $f(Q)$ theory, to simplify calculations, we usually use the coincident gauge, i.e., $\Gamma^\alpha_{\mu \nu} = 0$, so that the covariant derivative becomes a simple partial derivative. In this gauge:
\begin{equation}\label{gauge1}
Q_{\alpha \mu \nu} = \partial_\alpha \, g_{\mu \nu},
\end{equation}
and then the metric derivatives and non-zero components of the non-metricity tensor are:
\begin{equation}\label{met2}
\partial_0 \, g_{ij} = 2 a \dot{a} \delta_{ij},~~~ \partial_0 \, g_{00} = 0, ~~~~ Q_{0ij} = 2a \dot{a} \delta{ij},
\end{equation}
as a result, for the FLRW metric, we obtain the non-metricity scalar as
\begin{equation}\label{nms1}
Q = 6 H^2,
\end{equation}
where $H = \dot{a} / a$ is the Hubble parameter in which the dot represents the derivative with respect to cosmic time. Now we consider the universe as a perfect fluid, then the energy-momentum tensor for matter becomes
\begin{equation}\label{emt1}
T_{\mu \nu}\equiv-\frac{2}{\sqrt{-g}}\frac{\delta(\sqrt{-g})\mathcal{L}_m} {\delta g^{\mu \nu}} = (\rho + p) u_\mu u_\nu + p g_{\mu \nu},
\end{equation}
where $u_\mu = (-1, 0, 0, 0)$ is the four-velocity of the fluid and satisfies the condition $u_\mu u^\mu = -1$, \( \rho \) and \( p \) are the energy density and the pressure, respectively. In that case, we obtain the energy-momentum tensor in the FLRW metric as
\begin{equation}\label{emt2}
T^\mu_\nu = (-\rho, p, p, p).
\end{equation}

Now, by substituting the above obtained parameters into Eq. \eqref{Ein1}, we obtain the Friedmann equations for $f(Q)$ gravity in the following form
\begin{eqnarray}
&\kappa^2 \rho = 6 f_Q H^2 - \frac{1}{2} f, \label{Fried-1}\\
&\kappa^2 p = -2 f_Q \left(\dot{H}+3H^2\right) - 24 \dot{H} H^2 f_{QQ} + \frac{1}{2} f, \label{Fried-2}
\end{eqnarray}
where $f_{Q}=df / dQ$ and $f_{QQ}=d^2f / dQ^2$. We note that if $f = Q$, the standard gravitational model in general relativity is obtained. Now we clearly obtain the continuity equation of the present model as follows:
\begin{equation}\label{emt2}
\dot{\rho} + 3 H (\rho + p) = 0,
\end{equation}
where we consider the universe to consist of matter (dark matter and baryonic matter) and dark energy components with indices of $m$ and $de$ respectively as
\begin{eqnarray}
&\rho = \rho_{m} + \rho_{de}, \label{cont-1}\\
&p = p_{m} + p_{de}, \label{cont-2}
\end{eqnarray}
where the corresponding Friedmann equations rewrite as
\begin{eqnarray}
&\kappa^2 \rho_{de} = 6 f_Q H^2 - \frac{1}{2} f - \kappa^2 \rho_{m}, \label{Fried-3}\\
&\kappa^2 p_{de} = -2 f_Q \left(\dot{H}+3H^2\right) - 24 \dot{H} H^2 f_{QQ} + \frac{1}{2} f - \kappa^2 p_{m}, \label{Fried-4}
\end{eqnarray}
where we immediately obtain the continuity equations of the corresponding components with considering the interaction between them as follows:
\begin{eqnarray}
&\dot{\rho}_{m} + 3 H (1+\omega_{m}) \rho_{m}= \mathcal{Q}, \label{contdm}\\
&\dot{\rho}_{de} + 3 H (1+ \omega_{de}) \rho_{de} = - \mathcal{Q}, \label{contde}
\end{eqnarray}
where $\omega_{m} = p_{m} / \rho_{m}$ is the EoS of matter as this component is modeled as a cosmic effective fluid with the steady-state EoS and is not limited to standard cold dark matter, so, it is considered as a free phenomenological parameter. The quantity $\omega_{de} = p_{de} / \rho_{de}$ is EoS of dark energy, which characterizes the dynamic universe. Also, $\mathcal{Q}=3 b^2 H \rho_m$ term is introduced as an interaction between the matter and dark energy in which $b$ is a constant that one is related to energy strange between the components of the universe \cite{Amendola2000, Zimdahl2001, Wang2005, Wang2016}. Therefore, we can obviously obtain the solution to differential equation \eqref{contdm} as follows:
\begin{equation}\label{rhodm1}
\rho_{m} = \rho_{{m}_0} \left(\frac{a}{a_0}\right)^{-3 (1-b^2+\omega_{m})},
\end{equation}
where $\rho_{{m}_0}$  is the present value of $\rho_{m}$, and $a_0$ is the current value of the scale factor. Now, the EoS of dark energy is obtained by using Eqs. \eqref{Fried-3} and \eqref{Fried-4} as
\begin{equation}\label{EoS1}
\omega_{de} = -1 - \frac{2 f_Q \dot{H} + 24 \dot{H} H^2 f_{QQ} + \kappa^2 \rho_m + \kappa^2 p_m}{ 6 f_Q H^2 - \frac{1}{2} f - \kappa^2 \rho_m},
\end{equation}
so this equation provides our mathematical framework for determining whether our $f(Q)$ model can describe the observed behavior of the universe (accelerating expansion) and what future it predicts for the universe.


\section{HDE Model}\label{III}

The holographic principle, which has its roots in black hole thermodynamics and quantum gravity theories, states that the maximum information or independent degrees of freedom of a physical system in a region of space is proportional to the area of the boundary of that region, not to its volume \cite{tHooft1993,Susskind1995}. This principle states that a complete description of the physics in a three-dimensional volume can be encoded on the two-dimensional boundary surface of that volume. The application of this idea to cosmology has provided a natural motivation for introducing the holographic dark energy model, in which the energy density of the vacuum depends on an infrared cutoff length scale \cite{Cohen1999,Hsu2004,Li2004,Wang2017}.

According to the holographic principle, the entropy of a physical system of a given size \(L\) cannot exceed the entropy of a black hole of the same size. Consequently, the Bekenstein–Hawking bound for entropy is given by
\begin{equation}
S_{BH}=\frac{A}{4G}=8\pi^2M_p^2L^2,
\end{equation}
where \(A=4\pi L^2\) is the horizon area, \(G\) is the universal gravitational constant, and \(M_p^2=1/\kappa^2=(8\pi G)^{-1}\) is the reduced Planck mass.

On the other hand, if the total energy of a field theory effective in a volume of \(L^3\) exceeds the mass of a black hole of the same size, the system becomes gravitationally unstable and collapses, forming a black hole. Therefore, it is necessary to have the condition
\begin{equation}
L^3\rho_D\le M_p^2L,
\end{equation}
where sets an upper bound for the energy density in a region of a given size \(L\) \cite{Cohen1999}. Assuming saturation of this bound, the holographic dark energy density is defined as
\begin{equation}
\rho_D=3c^2M_p^2L^{-2}
=\frac{3c^2}{\kappa^2}L^{-2},
\end{equation}

where \(c\) is a dimensionless parameter that controls the degree to which the model deviates from the cosmological constant. \(L\) also represents the infrared cutoff of the theory, which, depending on the model, can be chosen as the future event horizon, the Hubble horizon, the apparent radius, or other cosmological scales. The appropriate choice of this length scale plays a decisive role in the dynamical behavior of dark energy and the expansion history of the universe.

In this paper, with the aim of investigating the behavior of holographic dark energy in the gravitational framework \(f(Q)\), the above holographic energy density is introduced as a dark energy component in the modified field equations. Then, by considering the interaction between dark matter and dark energy and adopting a suitable parameterization for the Hubble parameter, the corresponding gravitational function is reconstructed and the cosmological consequences and propagation of gravitational waves are studied.

\section{Correspondence between HDE and $f(Q)$ gravity}\label{IV}

Modified gravity theories, especially those based on the non-metric scalar $f(Q)$, offer a powerful approach to explaining dark energy without introducing a separate scalar fields such as quintessence, phantom, etc. In this framework, the field equations are modified by changes with respect to the metric, using the non-metric scalar $Q$ (defined from the non-metric scalar $Q$). The main goal is to find a functional form of $f(Q)$ that captures the observed cosmological behavior, in particular the accelerating expansion. This can be achieved analytically by applying the dark energy holographic principle as a constraint on the structure of $f(Q)$.

To make this correspondence, it is first assumed that the universe evolves according to a power law for the cosmological scale factor as \cite{Nojiri2007, Singh2024}
\begin{equation}\label{att0}
a(t) = a_0 \left(\frac{t}{t_0}\right)^n,
\end{equation}
where $n$ is a parameter dependent on the energy content (dark matter and dark energy) and represents the dominant type of matter at that time, and $a_0$ and $t_0$ are the present values of the scale factor and the universe age, respectively. This power law represents the dominance of a particular form of matter/energy (whose density parameters are encoded in $n$) at a particular time interval. An important motivation for choosing the power-law scale factor is its analysability. Unlike many phenomenological parameterizations, it allows for the exact reconstruction of the corresponding function $f(Q)$ so that we can calculate cosmological parameters, and allows for a fully analytical study of thermodynamic and cosmological properties. Using the relationship between the scale factor and the redshift parameter, i.e.
\begin{equation}\label{at1}
a(t) = \frac{a_0}{1+z},
\end{equation}
and the useful differential relation
\begin{equation}\label{ddt1}
\frac{d}{dt}=-H (1+z) \frac{d}{dz},
\end{equation}
we obtain the corresponding equations in the following form
\begin{eqnarray}
&H = \frac{n}{t}, \label{Hnt1}\\
&H_0 = \frac{n}{t_0}, \label{Hnt2}\\
&H(z) = H_0 (1+z)^\frac{1}{n}, \label{Hnt3}\\
&\dot{H} = -(1+z) H\, \partial_z H = - \frac{1}{n} H^2, \label{Hnt4}\\
&\rho_{m} = \rho_{{m}_0} (1+z)^{3 (1-b^2+\omega_{m})},\label{Hnt5}
\end{eqnarray}
where $H_0$ is the present Hubble parameter.

According to the holographic principle, the dark energy density $\rho_D$  that drives the accelerating expansion is holographically related to the area of the cosmological horizon. In this paper, the cosmological horizon is considered the Hubble horizon because it is related to the instantaneous expansion rate of the universe and is proportional to the inverse of the Hubble parameter ($L=H^{-1}$), i.e.,
\begin{equation}\label{SBH3}
\rho_D = \frac{3 c^2}{\kappa^2} H^2,
\end{equation}
where to strengthen the theoretical foundations of the present paper, we establish a correspondence between the dark energy density $\rho_{de}$ and the HDE density $\rho_D$. In this case, Eq. \eqref{Fried-3} yields
\begin{equation}\label{Fried3}
\kappa^2 \rho_{{m}_0}(1+z)^{3(1-b^2+\omega_{m})} + 3 c^2 H_0^2 (1+z)^{\frac{2}{n}} = 6 H_0^2 (1+z)^{\frac{2}{n}} f_Q - \frac{f}{2},
\end{equation}
where clearly gives the solution to the corresponding differential equation as follows:
\begin{equation}\label{fz1}
  f(z) = 6 c^2 H_0^2 (1+z)^{\frac{2}{n}} + \frac{2 \kappa^2 \rho_{m_0}}{3n(1-b^2+\omega_{m})-1} (1+z)^{3(1-b^2+\omega_{m})} + c_0 \sqrt{6 H_0^2} (1+z)^{\frac{1}{n}},
\end{equation}
where $c_0$ is an integral constant. Then, by using
\begin{equation}\label{zQ1}
1+z = \left(\frac{Q}{6 H_0^2}\right)^\frac{n}{2},
\end{equation}
immediately yields
\begin{equation}\label{fz2}
  f(Q) = c^2 Q + c_1 Q^{m} + c_0 \sqrt{Q},
\end{equation}
where $c_1=\frac{2  \kappa^2\,\rho_{m_0}}{6^{m} H_0^{2 m} (2 m-1)}$, $m=\frac{3}{2} n (1-b^2+\omega_{m})$. This fundamental relation determines the structure of the function $f(Q)$ in the field equations such that the dark energy correction effects are a binomial combination of linear and power terms. The linear term $Q$ ensures the standard gravitational field behavior, while the power term $Q^m$ directly reproduces the effects of dark energy coupled to the geometry via holography. Fine-tuning the constants based on current values of the Hubble parameters and matter density makes this model $f(Q)$ a powerful tool for modeling the dynamics of the universe. To continue the work and facilitate the corresponding calculations, we consider the integral constant value as $c_0 = 0$, hence function $f(Q)$ is considered as $ f(Q) = c^2 Q + c_1 Q^{m}$.

Now we can obtain the energy density and the pressure from Eqs. \eqref{Fried-3} and \eqref{Fried-4} for dark energy component in terms of the redshift parameter in the following form
\begin{eqnarray}
 & \rho_{de} = \frac{3 c^2}{\kappa^2} H_0^2 (1+z)^{\frac{2}{n}},\label{Fried31}\\
  & p_{de} = -b^2 \rho_{m_0} (1+z)^{3(1-b^2+\omega_{m})} - \frac{(3 n-2) c^2}{n \kappa^2} H_0^{2} (1+z)^{\frac{2}{n}},\label{Fried32}
\end{eqnarray}
also, the EoS for dark energy yields
\begin{equation}\label{EoS2}
\omega_{de} = -1 + \frac{2}{3 n} - \frac{b^2 \kappa^2 \rho_{m_0}}{3 c^2 H_0^2} (1+z)^{\frac{-2}{n}+3(1-b^2+\omega_m)},
\end{equation}
where this relationship shows that the EoS of dark energy depends on the components of matter, the interaction term and holographic. It should be noted that the above result has an interesting consequence. In this sense, if the universe is considered without the interaction factor ($b=0$), $\omega_{de} = -1 + \frac{2}{3 n}$, which means that the universe is stationary. Therefore, in order to have a dynamic universe, the existence of the interaction term between the dark parts of the universe that originates purely from phenomenology plays an essential role.

Therefore, cosmological parameters help us to understand the history of the universe from the Big Bang to the present. The deceleration parameter is another quantity that increases our knowledge about the formation of the universe and is expressed as follows:
\begin{equation}\label{q1}
  q = -1 - \frac{\dot{H}}{H^2},
\end{equation}
where by putting Eq. \eqref{Hnt4} into it we get $q = -1 + \frac{1}{n}$. We note that $q$ is not an independent quantity, but is determined directly from the power law ansatz itself. However, for the universe to be in an accelerating phase, the condition $q < 0$ must hold, in which case, $n > 1$ must hold. To this end, the condition ($n > 1$) is expected to be satisfied by the Hubble data constraint, which will be discussed in the next section.

In the next section, we will examine the fitting of the present model to observational data, including the Hubble data set.

\section{Observational constraints on the Hubble data set}\label{V}

In what follows, the observational aspects of the present model are analysed. In this paper, We only use the $H(z)$ data set according to Tab. \ref{table1} to obtain the best fit value for the parameter $n$. The contents of this table are in the range $0.07 \le z \le 2.36$, measured by galaxy differential age method and radial BAO size method. To optimally estimate the corresponding free parameter, the chi-square function was used as a fit criterion between the observational data and the present theoretical model in the following form
\begin{equation}\label{chi1}
\chi^2_{\min} = \sum_{i=1}^{46} \frac{(H_{\text{o}}(z_i) - H_{\text{t}}(z_i, H_0))^2}{\sigma^2_{i}(z_i)},
\end{equation}
where $H_o$, $H_t$, and $\sigma_i$ are the observed Hubble parameter values, the theoretical Hubble parameter values, and the observed standard error values, respectively. In that case, the chi-square function was optimized by minimizing it with respect to the free parameter $n$ for the power-law model \eqref{Hnt3}. 

\begin{table}[h]
\caption{The observational Hubble parameter data consist of 46 measurements obtained from cosmic chronometers and BAO observations.} 
\centering 
\begin{tabular}{||c | c | c | c | c || c | c | c | c | c|| c | c | c | c | c ||} 
\hline\hline 
No. & ~z~ & ~H(z)~ & $~\sigma_{i}~$ & ~Ref.~ &No. & ~z~ & ~H(z)~ & $~\sigma_{i}~$ & ~Ref.~ & No. & ~z~ & ~H(z)~ & $~\sigma_{i}~$ & ~Ref.~\\ [0.5ex] 
\hline 
1.   & 0.07 & 69.0 & 19.6 & \cite{Zhang2014}   & 17. & 0.4004 & 77 & 10.2 &  \cite{Moresco2016}   & 33. & 0.875 & 125.0 & 17.0 & \cite{Moresco2012}\\
2.   & 0.09 & 69 & 12.0 & \cite{Zhang2014}   & 18. & 0.4247 & 87.1 & 11.2 &  \cite{Moresco2016}   & 34. & 0.88 & 90.0 & 40.0 &  \cite{Stern2010}\\
3.   & 0.12 & 68.6 & 26.2 & \cite{Zhang2014}      & 19. & 0.43 & 86.5 & 3.7 &  \cite{Gaztanaga2009} & 35. & 0.9 & 117.0 & 23.0 &  \cite{Simon2005}\\
4.   & 0.17 & 83.0 & 8.0 &  \cite{Moresco2012}      & 20. & 0.44 & 82.6 & 7.8 &  \cite{Blake2012}  & 36. & 1.037 & 154.0 & 20.0 &  \cite{Moresco2012}\\
5.   & 0.179 & 75.0 & 4.0 & \cite{Moresco2012}     & 21. & 0.4497 & 92.8 & 12.9 &  \cite{Moresco2016} &  37. & 1.3 & 168.0 & 17.0 & \cite{Simon2005}\\
6.   & 0.1993 & 75.0 & 5.0 &  \cite{Moresco2012}  & 22. & 0.47 & 89.0 & 49.6 &  \cite{Ratsimbazafy2017}  & 38. & 1.363 & 160.0 & 33.6 &  \cite{Moresco2015}\\
7.   & 0.2 & 72.9 & 29.6 &  \cite{Zhang2014}  & 23. & 0.4783 & 80.9 & 9 &  \cite{Moresco2016} & 39. & 1.43 & 177.0 & 18.0 &  \cite{Simon2005}\\
8.   & 0.24 & 79.7 & 2.7 &  \cite{Gaztanaga2009}&  24. & 0.48 & 97 & 60 &  \cite{Stern2010}  &40. & 1.53 & 140.0 & 14.0 &  \cite{Simon2005}\\
9.   & 0.27 & 77.0 & 14.0 &  \cite{Zhang2014}      & 25. & 0.51 & 90.4 & 1.9 & \cite{Chuang2013}    & 41. & 1.75 & 202.0 & 40.0 &  \cite{Simon2005}\\
10. & 0.28 & 88.8 & 36.6 & \cite{Zhang2014}     & 26. & 0.57 & 96.8 & 3.4 &  \cite{Anderson2014}  & 42. & 1.965 & 186.5 & 50.4 &  \cite{Moresco2015}\\
11. & 0.31 & 78.17 & 6.74 &  \cite{Wang2017}        & 27. & 0.593 & 104 & 13 &  \cite{Moresco2012}  & 43. & 2.3 & 224 & 8.6 &  \cite{Busca2013} \\
12. & 0.35 & 82.7 & 8.4 &  \cite{Chuang2013}    & 28. & 0.6 & 87.9 & 6.1 &  \cite{Blake2012}  & 44. & 2.33 & 224 & 8 &  \cite{Bautista2017}\\
13. & 0.352 & 83 & 14 &  \cite{Zhang2014}  & 29. & 0.61 & 97.3 & 2.1 &  \cite{Alam2017} & 45. & 2.34 & 222 & 7 & \cite{Delubac2015}\\
14. & 0.38 & 81.5 & 1.9 &  \cite{Alam2017}     & 30. & 0.68 & 92 & 8 & \cite{Moresco2012} & 46. & 2.36 & 226 & 8 &  \cite{FontRibera2014}\\
15. & 0.3802 & 83 & 13.5 &  \cite{Moresco2012}  & 31. & 0.73 & 97.3 & 7 &  \cite{Blake2012} &  &  &  &  & \\
16. & 0.4 & 95 & 17 &  \cite{Alam2017}         & 32. & 0.781 & 105.0 & 12 &  \cite{Moresco2012} &  &  &  &  & \\
[1ex] 
\hline 
\end{tabular}
\label{table1} 
\end{table}

\begin{figure}[t]
\begin{centering}
\subfigure[]{\includegraphics[scale=0.35]{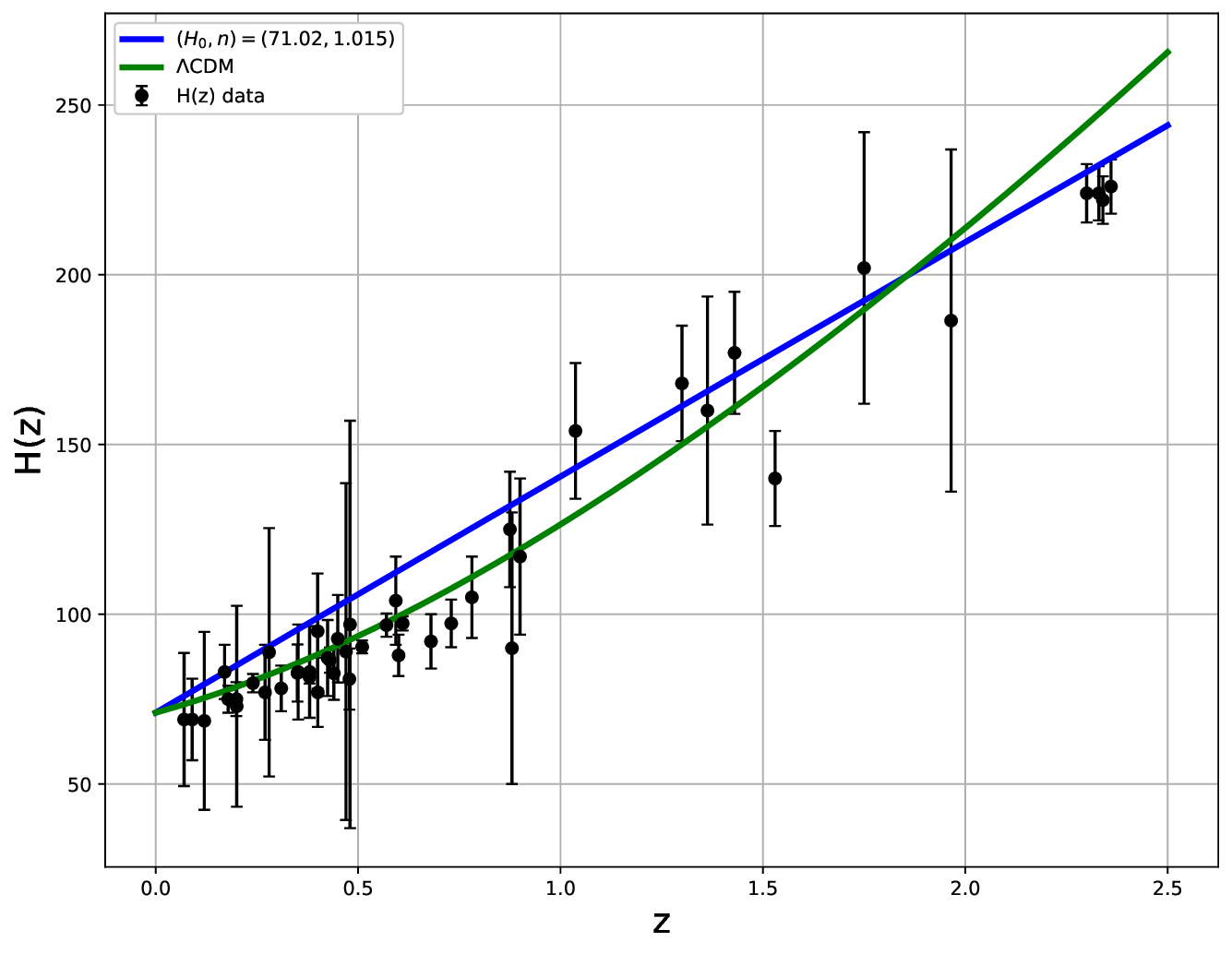}\label{H2zn2H0-a}}
\subfigure[]{\includegraphics[scale=0.4]{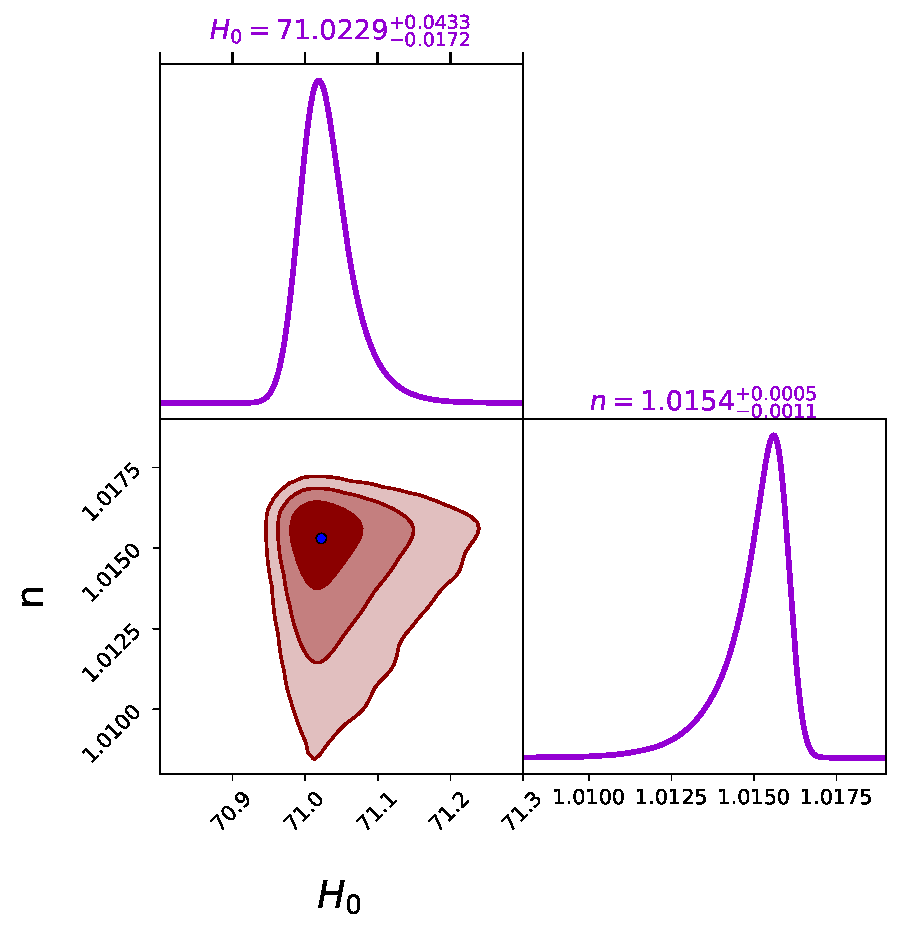}\label{H2zn2H0-b}}
\caption{The left panel (a): variations of $H(z)$ versus redshift for 46 Hubble data (bullets), best-fit model (blue line), and the $\Lambda CDM$ model (green line). The right panel (b): The contour plots for ($H_0, n$) at 1$\sigma$, 2$\sigma$, and 3$\sigma$ confidence levels.}
\label{H2zn2H0}
\end{centering}
\end{figure}

With these explanations, we plot the Hubble parameter versus redshift for three cases, such as the best fit of the current model, the $\Lambda \text{CDM}$ model, and 46 Hubble data set, as shown in Fig. \ref{H2zn2H0-a}. We also plot the best fit contour $H_0 = 71.0229^{+0.0433}_{-0.0172} ~{\rm km\,  s^{-1}  Mpc^{-1}}$ and $n = 1.0154^{+ 0.0005}_{- 0.0011}$ at three confidence levels of 1$\sigma$, 2$\sigma$, and 3$\sigma$ as shown in Fig. \ref{H2zn2H0-b}. 

In order to statistically constrain the free parameters of the model, we have used a set of 46 Hubble bounds \(H(z)\). From the Fig. \ref{H2zn2H0-a}, we see that the blue curve shows the behavior of our proposed model with the best-fit values, i.e., \((H0, n)\) = (71.0229, 1.0154). As can be seen, in the late universe \(z \lesssim 1.0\), our interaction model is in excellent agreement with the standard $\Lambda\text{CDM}$ model (green curve) and the observational data due to the dominant contribution of the holographic density term and the energy exchange in the dark sector. However, at higher redshifts ($z > 1.5$), our model curve shows a gentler slope than $\Lambda\text{CDM}$. This structural deviation is directly due to the $c_1 Q^m$ correction term in the gravitational action and non-gravitational interaction of the sources, because in the cosmic past, the non-metricity density $Q$ was larger and the modified gravity injects a different behavior into the model than the standard cosmological constant. The Fig. \ref{H2zn2H0-b} displays the smoothed two-dimensional probability distribution contours at the confidence levels of $1\sigma$, $2\sigma$, and $3\sigma$ in the phase space of $(H_0, n)$ along with the one-dimensional probability distribution functions. Statistical analysis of the optimization process yields the following optimal values at the 68\% confidence level ($1\sigma$): $H_0 = 71.0229_{-0.0172}^{+0.0433} ~{\rm km\,  s^{-1}  Mpc^{-1}}$, $n = 1.0154_{-0.0011}^{+0.0005}$. The one-dimensional distribution curve for the geometric power parameter $n$ is very symmetric and narrow, indicating the rigidity and strong constraint of this parameter by the data of cosmic sources. From a physical point of view, the obtained value $n = 1.0154$ (very close to unity) shows that the Hubble ansatz exponent has a subtle deviation from the linear state, which directly reflects the coupling of the nonstandard term $Q^m$ in the Friedmann field equations. Moreover, the geometric orientation of the two-dimensional contour ellipses reveals a slight positive correlation between $H_0$ and $n$, this means that the interaction term in the model allows a small increase in the power of the cosmic evolution $n$ to statistically overlap with the adjustment of the current expansion rate $H_0$.

One of the most striking results of this statistical analysis is the optimal value of the Hubble constant ($H_0 \approx 71.02 \text{ km/s/Mpc}$). In modern cosmology, the standard $\Lambda\text{CDM}$ model based on the cosmic microwave background (Planck CMB) data reports a low value for the Hubble constant as $H_0 = 67.4 \pm 0.5 ~{\rm km\,  s^{-1}  Mpc^{-1}}$ \cite{Planck2020}, while local astrophysical measurements (such as those of the SHOES team) show higher value of $H_0 = 73.04 \pm 1.04 ~{\rm km\,  s^{-1}  Mpc^{-1}}$ \cite{Riess2022, Riess2024}. This contradiction is known as the "Hubble tension". The model presented in this work effectively moderates this tension. In standard models of HDE with Hubble cutoff ($L=H^{-1}$), if there is no interaction, the dark energy EoS is preferred as an invariant constant that is unable to resolve the crisis. However, in our model, the presence of active interaction between the geometric gravity terms $f(Q)$ (via the analytically derived function) and the holographic density creates a cycle of energy exchange that re-adjusts the late accelerated phase to a higher local expansion rate ($H_0 \approx 71.02$) without violating the $H(z)$ data in the transition to high redshifts. This achievement makes the above interaction modified gravity model a fully physical and competitive alternative to the standard cosmological front.

The surprising result of the values obtained from the above fitting is that the age of the universe is not obtained without integrating the dynamical equations, but can be calculated directly from the power-law form of the Hubble parameter by Eq. \eqref{Hnt2}. In this case, the current age of the universe is calculated with the fitted values equal to \(t_0=13.98^{+0.010}_{-0.015}\, {\rm Gyr}\). This value is in very good agreement with the new observational estimates. In particular, measurements based on the cosmic microwave background radiation, type Ia supernovae and baryonic acoustic oscillations report an age of the universe of about \(13.8~{\rm Gyr}\). The closeness of the value obtained in the present model to this observational range indicates that the reconstruction performed in the gravitational framework \(f(Q)\) with an error of 1.3\% describes the time scale of the evolution of the universe with acceptable accuracy.

\begin{figure}[h]
\begin{centering}
\includegraphics[scale=0.35]{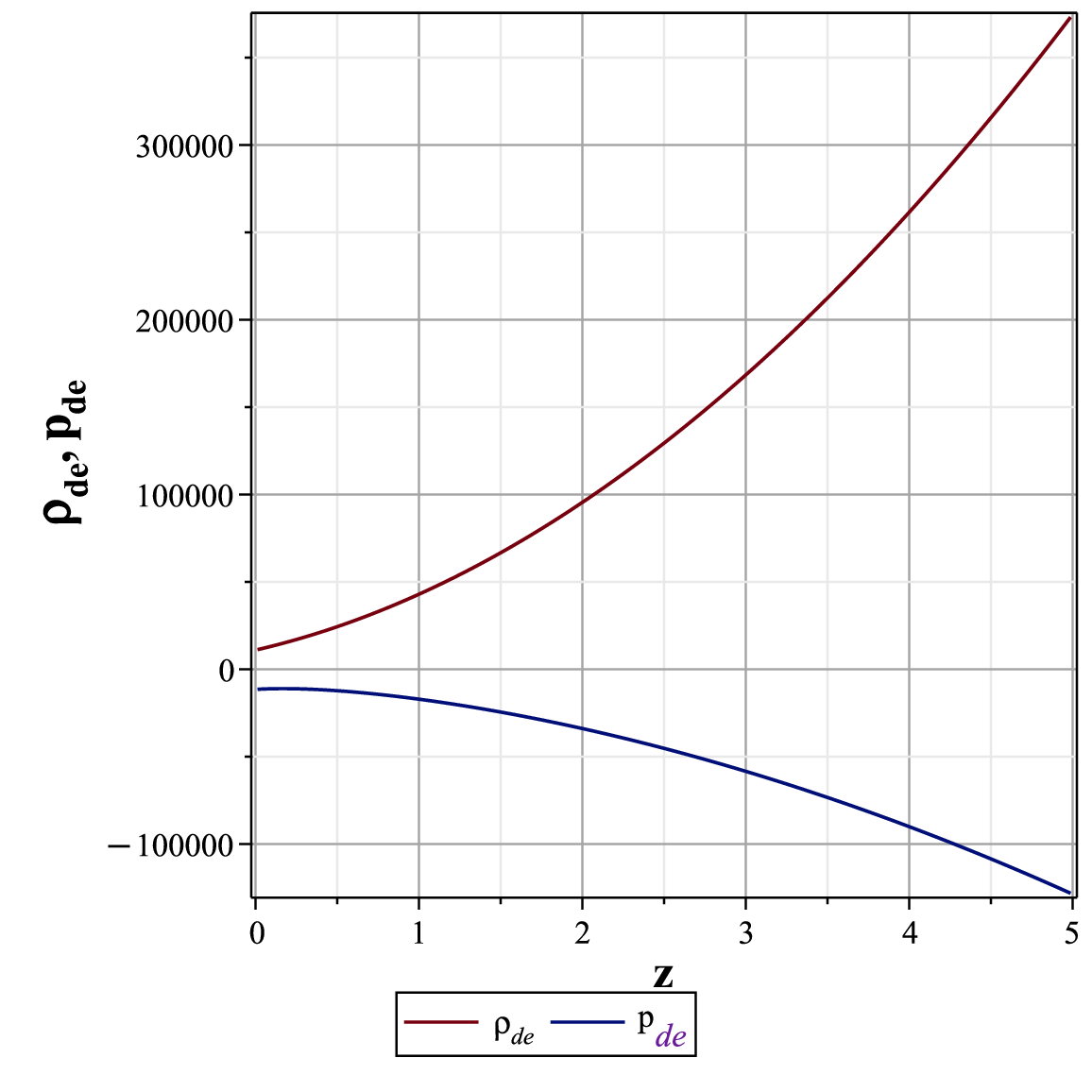}
\caption{Evolution of the energy density $\rho_{de}$ (solid red line) and pressure $p_{de}$ (solid blue line) of dark energy as a function of redshift $z$. The trajectories are plotted utilizing the best-fit and benchmark model parameters as $n = 1.0154$, $\omega_m = 0.01$, $\rho_{m_0} = 5000$, $c = 0.85$, and $b = 1.25$.}
\label{fig-rhop2z}
\end{centering}
\end{figure}

In Fig. \ref{fig-rhop2z}, in order to better understand the thermodynamic and hydrodynamic properties of the interaction modified gravity model $f(Q)$ coupled with HDE, the time evolution of the energy density $\rho_{de}$ and pressure $p_{de}$ of the dark energy is investigated in terms of redshift $z$. According to the findings of the diagram, the HDE density ($\rho_{de}$) is a completely ascending and smooth function of redshift, so that in the early universe (high redshifts, $z \to 5$) it has a much higher value and decreases gently as it approaches the present time ($z \to 0$). This behavior is fully consistent with the holographic principles based on the Hubble horizon $L=H^{-1}$, since as the universe expands and the Hubble rate decreases, the holographic density decreases correspondingly. On the other hand, the dark energy pressure curve \(p_{de}\) remains in the negative region \(p_{de} < 0\) throughout the entire cosmic evolution. This negative pressure is a crucial component to justify repulsive gravity and, consequently, to reconstruct the current accelerated phase of the universe. The important point from the point of view of the geometry of $f(Q)$ is that as we move from the cosmic past to the present \(z = 0\), the absolute value of the negative pressure decreases, but it still plays a dominant role in competing with the density of matter. This nonlinear evolution is directly due to the continuous energy exchange between the dark matter sector and modified gravity by the interaction term $b = 1.25$, which prevents the sudden decay of the energy density in the late universe.

\begin{figure}[h]
\begin{centering}
\includegraphics[scale=0.35]{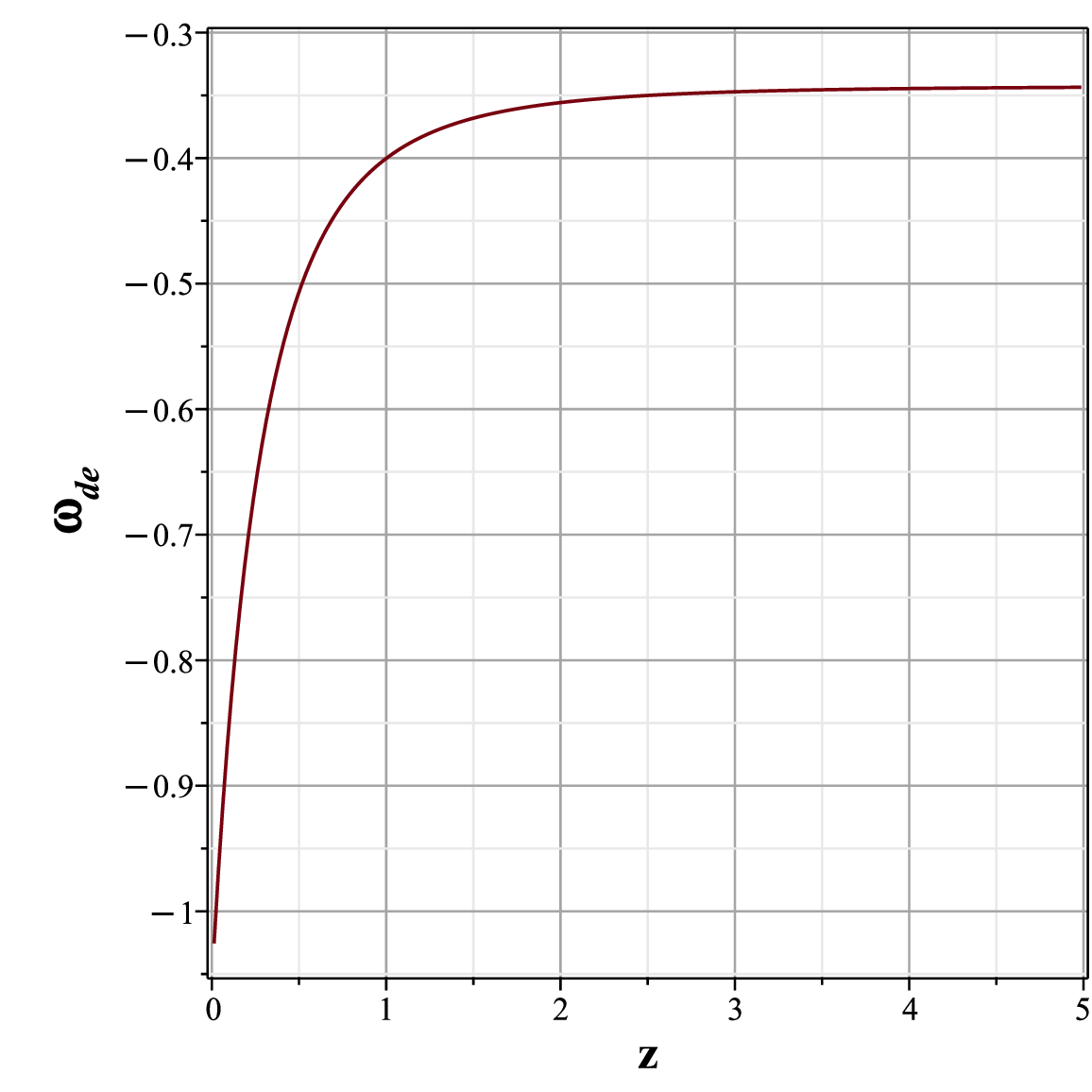}
\caption{Evolution of the dark energy EoS parameter as a function of redshift $z$. The trajectories are plotted utilizing the best-fit and benchmark model parameters as $n = 1.0154$, $\omega_m = 0.01$, $\rho_{m_0} = 5000$, $c = 0.85$, and $b = 1.25$.}
\label{fig-omega2z}
\end{centering}
\end{figure}

The Fig. \ref{fig-omega2z} shows the dynamic evolution of the dark energy EoS parameter $\omega_{de}$ with redshift, which is a key measure for determining the evolutionary behavior of the dark energy contribution. By tuning the model parameters to the specified values $n = 1.0154$, $\omega_m = 0.01$, $\rho_{m_0} = 5000$, $c = 0.85$, and $b = 1.25$, some very interesting results are obtained:

1. Physical justification of the values: We note that the value $\omega_m = 0.01$ (baryonic/dark matter EoS) is small, but considering it together with the large coupling parameter $b = 1.25$ shows that the model requires a relatively strong coupling in the dark sector to reach the cosmic transition.

2. Early universe behavior (Quintessence-like phase): At high redshifts ($z \ge 2$), the EoS parameter is stabilized in a range close to $\omega_{de} \approx -0.34$. This negative value indicates that even in past eras when matter dominated the universe, the geometric modified gravity $f(Q)$, along with the holographic structure, preserved the dark energy seed as a component with a mild negative pressure (similar to quintessence fields).

3. Dynamical transition and phantom crossing: As the redshift decreases and moves towards the present time ($z \to 0$), the $\omega_{de}$ curve experiences a sharp and steep downward slope. At the present time ($z = 0$), the EoS parameter approaches exactly the boundary of the cosmological constant line, $\omega_{de} = -1$, and even exhibits a slight "ghostly" or phantom behavior ($\omega_{de} \lesssim -1$). This dynamical transition from the quintessence phase to the phantom phase (Quintom behavior) is one of the most striking features of the interaction modified gravity models. 

In standard models of HDE with Hubble horizon cutoff, without the presence of interaction, the EoS parameter will always have a constant value ($\omega_{de} = -1 + \frac{2}{3 n}$) that is unable to cross the boundary $\omega_{de} = -1$ and practically fails to reconstruct the current acceleration of the universe. However, in our model, the presence of the interaction coupling term $b = 1.25$ in combination with the power-law index $n = 1.0154$ acts as a dynamic driving engine and, by transferring energy from the matter part to the holographic part, pushes the EoS parameter towards more negative values at the present time. This behavior is fully consistent with the latest cosmological observational data and is a testament to the structural success of the modified $f(Q)$ model presented in this paper.

In addition to examining the EoS parameter mentioned above, the deceleration parameter can play a complementary role in understanding the dynamics of dark energy within the framework of modern cosmology. The deceleration parameter in this model is obtained from the relation \(q=-1+1/n\). By choosing \(n=1.0154\), a constant value \(q\simeq-0.015\) is obtained, which indicates an accelerated phase in the expansion of the universe. It should be noted that the constancy of the deceleration parameter is due to the choice of a power-law ansatz for the Hubble parameter in the model reconstruction process and is considered as a limitation of the present reconstruction method. Therefore, the lack of observation of the transition between the deceleration and acceleration phases is a direct consequence of the choice of a power-law form for \(H(z)\) and not an intrinsic result of the \(f(Q)\) theory or the interaction between matter and dark energy.

\begin{figure}[h]
\begin{centering}
\includegraphics[scale=0.35]{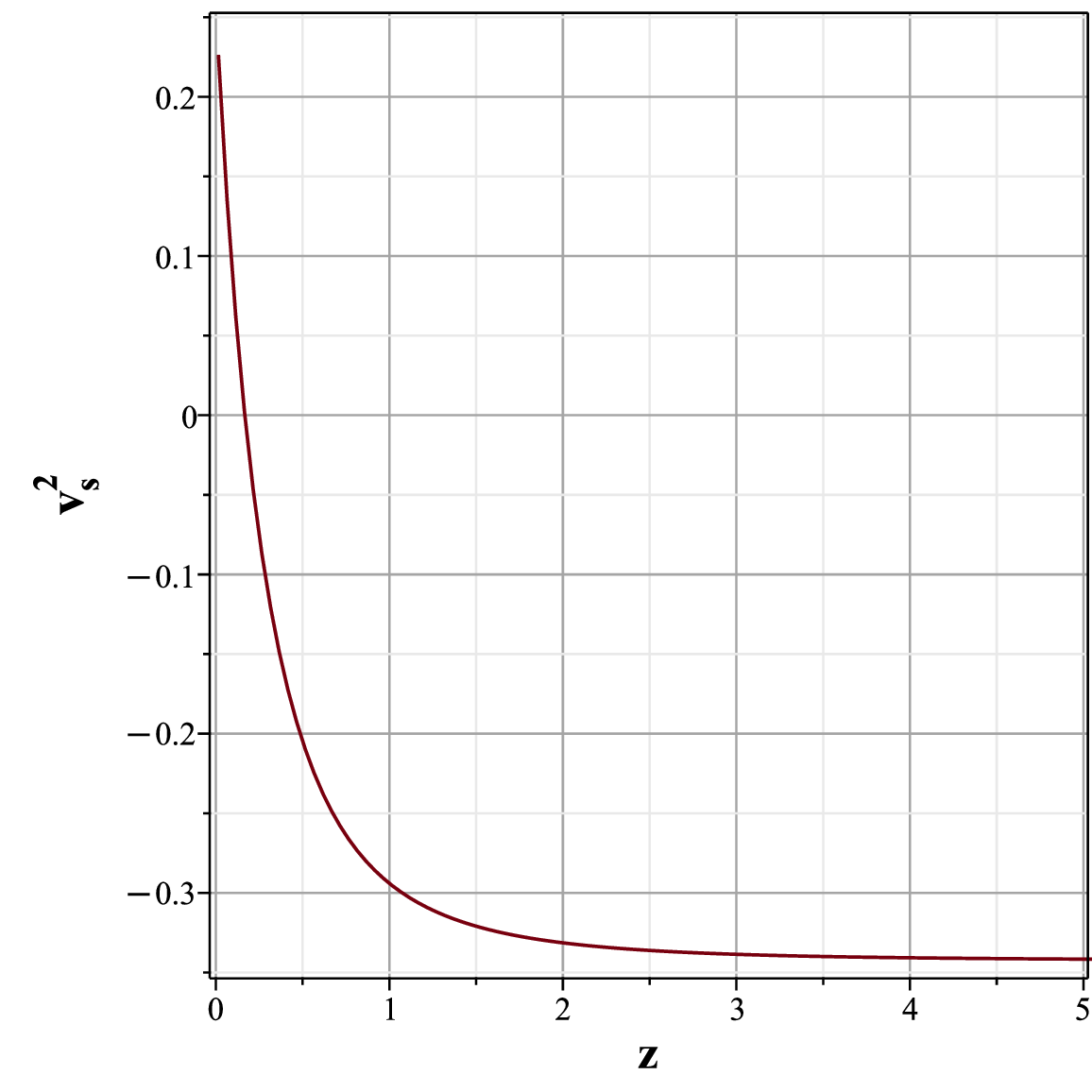}
\caption{The cosmic evolution of the squared speed of sound of the dark energy fluid versus redshift $z$. The trajectories are plotted utilizing the best-fit and benchmark model parameters as $n = 1.0154$, $\omega_m = 0.01$, $\rho_{m_0} = 5000$, $c = 0.85$, and $b = 1.25$.}
\label{fig-vs2z}
\end{centering}
\end{figure}

The analysis of the squared speed of sound $v_s^2$ is one of the most crucial parts of any cosmological investigation to ensure that our model does not suffer from physical instabilities such as explosive growth of density perturbations or non-standard causal behavior. For this purpose, the classical squared speed of sound parameter $v_s^2$ defines the propagation rate of first-order density perturbations in a cosmic fluid and is formulated as the partial derivative of pressure with respect to energy density as follows:
\begin{equation}\label{vs1}
v_s^2 = \frac{\partial p_{de}}{\partial\rho_{de}} = \frac{p'_{de}}{\rho'_{de}},
\end{equation}
where prime $(')$ is the derivative with respect to redshift $z$. From a physical point of view, and according to the Landau-Lifshitz hydrodynamic stability criterion, there are two basic conditions for the stability of a model:

1. Causality condition: $v_s^2 \le 1$ must hold so that the speed of propagation of disturbances does not exceed the speed of light.

2. Classical stability condition: $v_s^2 \ge 0$ must hold. If $v_s^2 < 0$, the frequency of local disturbances becomes imaginary ($\omega^2 = v_s^2 k^2 < 0$) and the disturbances grow exponentially instead of oscillating, leading to sudden collapse or the creation of physical instabilities in the fluid structure (called Laplace instability).

After examining the EoS parameter, it is of particular importance to evaluate the stability of the classical modified gravity model $f(Q)$ coupled with the HDE under small-scale thermal perturbations. To this end, we have plotted the dynamic evolution of the squared speed of sound $v_s^2$ in terms of redshift $z$ in Fig. \ref{fig-vs2z}. The geometric analysis of the $v_s^2$ evolution diagram brings profound physical achievements:

1. Stability in the late universe and the present ($z \to 0$): The most important finding of this diagram is the behavior of the $v_s^2$ in the late accelerated phase of the universe and the present ($0 \le z \le 0.25$). As can be seen, as we approach the present time ($z \to 0$), the value of $v_s^2$ of the holographic fluid becomes completely positive and falls within the causally permissible interval $0 \le v_s^2 \le 1$ (it reaches a value of about $0.22$ at the present time). This result shows that our interaction model is completely hydrodynamically stable in the current accelerated phase of the universe, and density perturbations in it do not undergo irrational explosive growth or divergence.

2. Instability at high redshifts ($z > 0.25$): As we move towards the cosmic past and higher values of redshift ($z \gtrsim 0.25$), the $v_s^2$ curve enters the negative region and stabilizes at a certain negative value ($v_s^2 \approx -0.34$). The negativity of the $v_s^2$ at high redshifts indicates the existence of a structural instability (Laplace instability) in the past of the universe for the dark energy component.

In the framework of interaction-modified gravitational cosmology, this instability in the past is not only not a defect, but is considered to be a desirable and physical behavior. Since in the transition to high redshifts ( $z > 1$), dark matter must be the dominant component of the universe for large-scale structures (clusters and galaxies) to form under the gravitational regime, the instability of the HDE component ( $v_s^2 < 0$) prevents the accumulation and clustering of dark energy itself at that time. In other words, this cosmic past instability ensures that dark energy behaves uniformly and without structure during the matter-dominated era and only at later times ( $z < 0.25$) does it enter a stable phase ( $v_s^2 > 0$) and take control of the dynamics of the universe to create the current positive acceleration. This dynamical stability transition is another confirmation of the consistency of the $f(Q) = c^2 Q + c_1 Q^m$ structure with standard observational cosmology.

In the following, in order to fully describe the present model, we will budget the energy contribution of the components of the current universe using the first Friedmann equation \eqref{Fried-3}. We note that in the framework of \(f(Q)\) gravity, the Friedmann equations are modified compared to the standard case of general relativity. Therefore, the definition of the critical energy density must also be consistent with the new structure of the field equations. In that case, the parameters of the relative densities of the matter and dark energy components of the universe are defined as follows, respectively:
\begin{eqnarray}
\Omega_m^{(\text{mod})} = \frac{\rho_m}{\rho_{\rm cr}^{(\text{mod})}}, \label{rhoc3-1}\\
\Omega_{de}^{(\text{mod})} = \frac{\rho_{de}}{\rho_{\rm cr}^{(\text{mod})}},\label{rhoc3-1}
\end{eqnarray}
where $\rho_{\rm cr}^{(mod)}=\frac{1}{\kappa^2}\left(6f_QH^2-\frac{f}{2}\right)$ is the effective critical density in \(f(Q)\) gravity. Note that in the general relativity limit, i.e. when $f(Q)=Q$ we have $\rho_{\rm cr}^{(\text{mod})}=\frac{3H^2}{\kappa^2}$ which is exactly the standard critical density in \(\Lambda\text{CDM}\) cosmology. The present definition is therefore a natural and consistent generalization of the conventional critical density to the gravitational framework \(f(Q)\). Using this definition, the current values of the relative density parameters with the specified values $\omega_m = 0.01$, $\rho_{m_0} = 5000$, $c = 0.85$, and $b = 1.25$ are obtained as
\begin{equation}\label{Ommde0}
\Omega_{m0}=0.314,
\qquad
\Omega_{de0}=0.686,
\end{equation}
where indicates that about 31\% of the effective energy budget of the current universe is allocated to matter and about 69\% to dark energy. This qualitative behavior is consistent with astronomical observations, including the results of the Planck mission (2018) and the $\Lambda\text{CDM}$ model, and shows that the holographic interaction model in the gravitational framework \(f(Q)\) is able to reproduce the energy composition of the present universe with reasonable accuracy. Furthermore, it follows directly from the above definition that
\begin{equation}\label{Ommde01}
\Omega_m+\Omega_{de}=1,
\end{equation}
where expresses the internal consistency of the model and the preservation of the flatness condition of the universe at the background level.

The overall results show that the studied holographic interaction model is capable of reproducing the observed contribution of matter and dark energy in the present universe, as well as describing the age of the universe and the reasonable behavior of the dark energy EoS parameter with appropriate accuracy. These features indicate the favorable compatibility of the model at the level of the cosmological background and provide a suitable platform for more detailed investigations, including the analysis of gravitational waves in the gravity framework \(f(Q)\). Therefore, in the next section, we will examine the gravitational wave dynamics for the present model.

\section{Gravitational Wave Dynamics in the Cosmological Background}\label{VI}

In this section, we discuss the dynamics of gravitational waves in the cosmological background. In this case, gravitational waves are translational fluctuations in the curvature of space-time that travel at the speed of light and arise from linearized wave-like solutions to Einstein's field equations. These waves carry information from strong gravitational phenomena such as black hole mergers, neutron stars, and also from the initial conditions of the universe during inflation. However, on cosmological scales, gravitational waves play an important role in the transformation of vacuum energy, the formation of the anisotropy of the cosmic microwave background (CMB), and the transmission of information from the inflationary era.

As discussed in the earlier sections of this paper, the \( f(Q) \) gravity is considered to be one of the simplest geometric generalizations of general relativity, in which the dynamics of gravity arises not from curvature but from a non-metricity scalar \( Q \). This framework is able to reproduce the accelerating expansion of the universe without the need to introduce explicit dark energy. The calculation of the GW equation in this theory has a twofold motivation: (1) from a theoretical point of view, investigating the behavior of tensor perturbations on a homogeneous and isotropic FRW background provides a tool for studying the dynamical stability of the theory at the linear order. (2) from an observational point of view, gravitational waves carry direct information about the fundamental structure of the gravitational field, hence deriving the wave equation in \( f(Q) \) allows investigating possible variations in the damping, velocity, and amplitude of propagation relative to the predictions of GR. Therefore, \( f(Q) \) theory is not simply an algebraic modification in the field equations, but changes the dynamics of gravitational waves themselves. These changes, although they may be small on a large cosmic scale, appear as detectable observational signatures in high-energy phenomena such as black hole or neutron star mergers.

To study the dynamics of gravitational waves in the context of cosmology, we describe the background space-time with the FLRW metric \eqref{met1}. Therefore, in the framework of linear perturbation theory, gravitational waves enter the metric as a transverse and traceless perturbation as follows:
\begin{equation}\label{met2}
  ds^2 = -dt^2 + a^2(t) \left( \delta_{ij} + h_{ij}(t, \vec{x}) \right) dx^i dx^j,
\end{equation}
provided that it satisfies the following transverse and traceless conditions, respectively, as
\begin{eqnarray}\label{transcon1}
  & \partial^i h_{ij} = 0, \label{transcon11}\\
  & h^i_i = 0, \label{transcon12}
\end{eqnarray}
so that these two conditions guarantee that only two physical degrees of freedom (two independent polarizations of gravitational waves) remain.

Now we expand $Q$ and $\sqrt{-g}$ from Eqs. \eqref{met2} and \eqref{nms1}, respectively, to the second order in $h_{ij}$ and the spatial and temporal derivatives of \(h_{ij}\) as
\begin{eqnarray}\label{sqrtg1}
  &\sqrt{-g} = a(t)^3 \left( 1 - \frac{1}{4} h_{ij} h^{ij} + \mathcal{O}(h^3) \right), \label{sqrtg11}\\
  &\delta Q = \frac{1}{4} \left[ \dot{h}_{ij}\dot{h}^{ij} - \frac{1}{a^2}(\partial_k h_{ij})(\partial^k h^{ij}) \right].\label{sqrtg12}
\end{eqnarray}

The second-order expansion of the function (with the zeroth and first-order terms that are constant or correspond to the background) is as follows:
\begin{equation}\label{GW1}
  S_{GW} = \frac{1}{8} \int d^3x\, dt\, a^3 f_Q \left[ \dot{h}_{ij}\dot{h}^{ij} - \frac{1}{a^2} (\partial_k h_{ij})(\partial^k h^{ij}) \right],
\end{equation}
then the gravitational wave equation can be derived from the principle of least action using the Euler–Lagrange equation for the field \(h_{ij}\) as
\begin{equation}\label{GW2}
  \ddot{h}_{ij} + \left(3H + \frac{\dot{f_Q}}{f_Q}\right)\dot{h}_{ij}- \frac{1}{a^2}\nabla^2 h_{ij} = 0.
\end{equation}

If we consider gravitational waves as an oscillatory mode with a certain spatial wave number $k$, which is decomposed as $h_{ij}(t, \vec{x}) \propto \frac{\eta(t)}{a(t)^2} e^{i \vec{k}\cdot \vec{x}}$,the governing equation for the above tensor equation becomes an ordinary differential equation for the oscillation amplitude $\eta(t)/a(t)^2$ as follows:
\begin{equation}\label{GW3}
  \ddot{\eta}(t) - \left(H - \frac{\dot{f}_Q}{f_Q}\right)\dot{\eta}(t) + \left(\frac{k^2}{a^2} - \frac{2 \ddot{a}}{a} - 2 H \frac{\dot{f}_Q}{f_Q}\right) \eta(t) = 0,
\end{equation}
this equation is the heart of the dynamics of gravitational waves and describes the dynamics of $\eta(t)$ fluctuations in the cosmological theory of $f(Q)$ gravity. This equation is a second-order linear differential equation with time-dependent coefficients, in which the term \( \ddot{\eta}(t) \) represents the acceleration of the disturbance, the damping term \(-\left(H - \frac{\dot{f}_Q}{f_Q}\right)\dot{\eta}(t)\) expresses the effect of the expansion of the universe \( H \) and the dynamics of the function \( f_Q \) on the reduction of the amplitude of the oscillations, and the coefficient term $\eta(t)$ includes several important physical effects: the spatial part \(\frac{k^2}{a^2}\), where \(k\) is the wave number and is related to the spatial structure of the oscillation, the geometric part \(-\frac{2\ddot{a}}{a}\) resulting from the acceleration of the expansion of space-time, and the term \(-2H\frac{\dot{f}_Q}{f_Q}\) which represents the role of dynamical modifications resulting from the \(f(Q)\) theory. In general, this equation describes the behavior of a damped wave in an expanding medium with modified dynamics.

In what follows, by using Eqs. \eqref{at1} and \eqref{ddt1}, we write down Eq. \eqref{GW3} in terms of the redshift parameter as
\begin{equation}\label{GW4}
  \eta''(z) + \left(-\frac{a^2 \ddot{a}}{\dot{a}^2} + 3 a - \frac{a^2}{\dot{a}} \frac{\dot{f}_Q}{f_Q}\right) \eta'(z) +  \left(\frac{a^2 k^2}{\dot{a}^2} - \frac{2 a^3 \ddot{a}}{\dot{a}^2} - \frac{2 a^3}{\dot{a}} \frac{\dot{f}_Q}{f_Q}\right) \eta(z) = 0,
\end{equation}
by using Eq. \eqref{fz2}, we immediately obtain the present equation in the following form
\begin{eqnarray}\label{GW5}
 & \eta''(z) + \frac{1}{n} \left(\frac{1+2 n}{1+z} + \frac{2 c_1 m (m-1) 6^{m-1} H_0^{2 (m-1)} (1+z)^{\frac{2(m-1)-n}{n}}}{c^2 + m c_1 6^{m-1} H_0^{2(m-1)} (1+z)^{\frac{2(m-1)}{n}}}\right) \eta'(z) \notag \\
  &+  \frac{1}{n} \left(\frac{n k^2}{H_0^2 (1+z)^{\frac{2}{n}}} + \frac{2 (1-n)}{(1+z)^2} + \frac{4 c_1 m (m-1) 6^{m-1} H_0^{2(m-1)} (1+z)^{\frac{2(m-n-1)}{n}}}{c^2 + c_1 m 6^{m-1} H_0^{2(m-1)} (1+z)^{\frac{2(m-1)}{n}}}\right) \eta(z) = 0,
\end{eqnarray}
where we arrive at a second-order differential equation of the function $\eta(z)$. Based on the theoretical foundations developed in relations \eqref{GW1} to \eqref{GW5}, the auxiliary field quantity $\eta$ is simply a mathematical variable in isomorphic time, while the real and observable amplitude of gravitational waves, which is directly affected by the expansion of the universe, is proportional to the covariant variable $h_{ij} \propto \frac{\eta}{a^2}$. Therefore, by numerically solving the governing differential equation \eqref{GW5}, we have simulated the exact evolution of the quantity $\frac{\eta}{a^2}$ over a wide cosmic interval from the early matter-dominant geometry era ($z = 150$) to the present ($z \to 0$). To make the calculations more realistic and avoid the artificial in-phase regime, the initial conditions for integration in the distant past are set to $\eta(150) = 0.9$ and $\eta'(150) = 0.1$. This initial heterogeneity allows different modes to exhibit distinct amplitude behaviors with respect to their structural frequency as shown in Fig. 5.

\begin{figure}[h]
\begin{centering}
\includegraphics[scale=0.35]{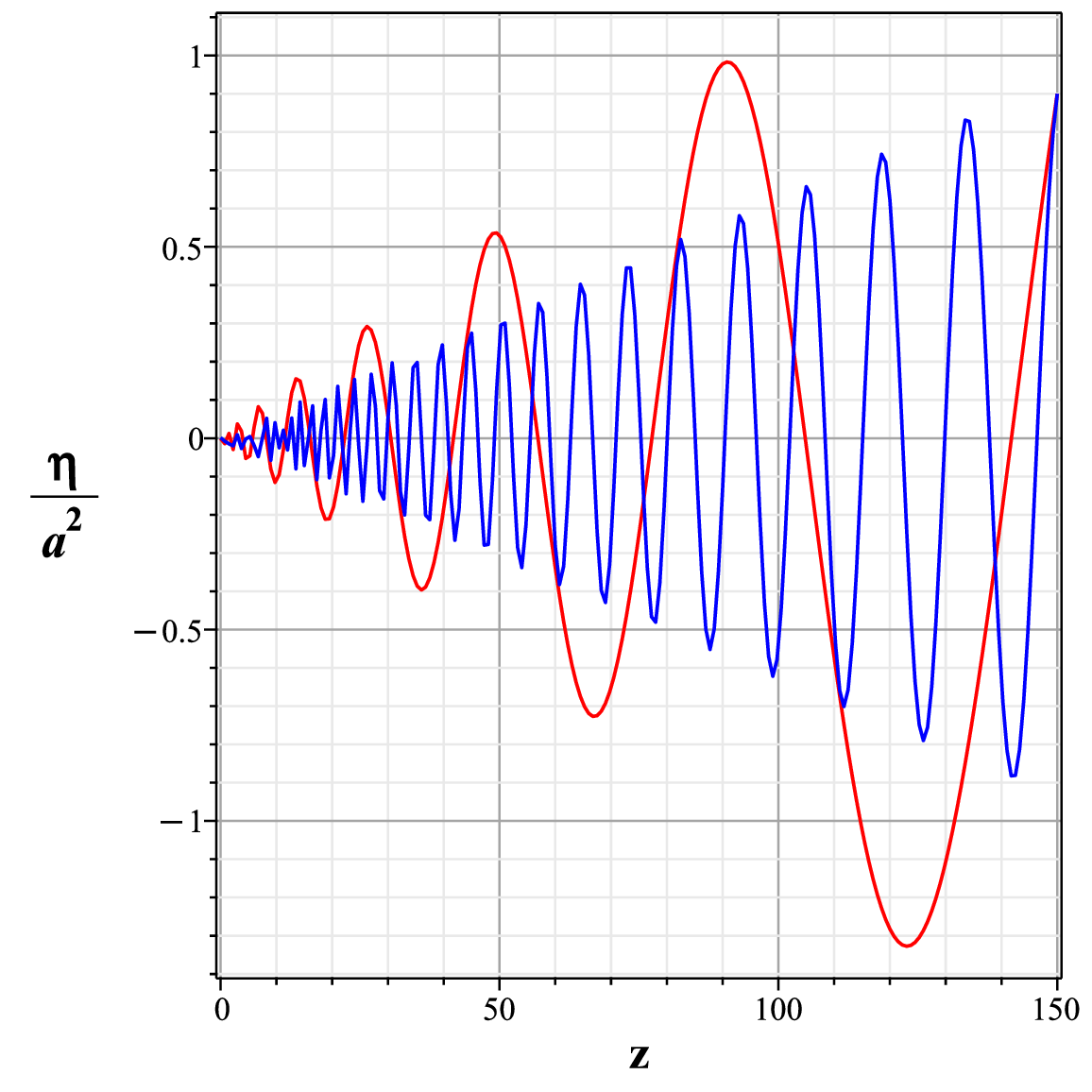}
\caption{The cosmic evolution of the gravitational wave amplitude $\eta/a^2$ as a function of $z$. The model parameters are fixed based on the optimal observational values: $n = 1.0154$, $\omega_m = 0.01$, $\rho_{m_0} = 5000$, $c = 0.85$ and $b = 1.25$ and the initial conditions are set to physically distinct $\eta(150) = 0.9$ and $\eta'(150) = 0.1$ with modes $k=700$ (red) and $k=3500$ (blue).}
\label{fig-eta2z}
\end{centering}
\end{figure}

Fig. \ref{fig-eta2z} shows the oscillatory behavior of the quantity $\eta/a^2$ in terms of redshift $z$ for two different wave numbers $k=700$ (red curve) and $k=3500$ (blue curve). The calculations were performed using the fixed parameters of the model as $n = 1.0154$, $\omega_m = 0.01$, $\rho_{m_0} = 5000$, $c = 0.85$ and $b = 1.25$, the resulting cosmic expansion is consistent with the observational data. The geometric analysis of the oscillation paths in Fig. \ref{fig-eta2z} reveals the following physical achievements:

1. Standard cosmological damping mechanism and energy dilution:
The most striking feature of the diagram is the systematic and uniform decrease in the physical amplitude of the wave ($\frac{\eta}{a^2}$) for both 700 and 3500 wave modes as one moves from the distant past to the present ($z = 150 \to 0$). From a physical point of view, as the fabric of spacetime expands and the scale factor of the universe increases ($a$), the energy density of the background gravitational waves becomes strongly diluted and the envelope of the oscillations undergoes a permanent decrease. This amplitude attenuation is fully consistent with the standard principles of tensor cosmology and the physical stability of the model.

2. Dynamic frequency enhancement and scale dependence:
The application of dynamic initial conditions has led to a very beautiful amplitude separation between different scales. The mode with a smaller wave number ($k=700$) as the main envelope experiences larger amplitudes in the evolution range, while the mode with a larger wave number ($k=3500$) exhibits much faster and more compact oscillations due to its higher frequency. As the present time approaches ($z \to 0$), the density of oscillations increases in terms of redshift, which is due to the dependence of the physical frequency of the wave on the scale factor ($\omega \propto k/a$) in an expanding medium.

3. Subtle geometric deviations from general relativity: Although tensor modes in the early universe ($z > 50$) reproduce the oscillatory regime of general relativity damping, upon entering the late universe ($z < 1$) and the onset of HDE dominance, the damping rate undergoes subtle changes. In standard gravity, the wave damping is solely influenced by the Hubble friction term (the term $-H \dot{\eta}$ in Eq. \eqref{GW3}). However, in our proposed model, the coefficient of the first derivative term in Eq. \eqref{GW5} contains the coupling terms $f_Q$ and the cosmic active interaction parameter ($b=1.25$). These geometric modifications reset the gravitational wave decay rate at later times. This slight deviation from the standard $\Lambda\text{CDM}$ model creates a unique observational fingerprint for the scenario of holographic $f(Q)$ gravity, which could be tested and detected in the future by sensitive third-generation gravitational wave detectors (such as the LISA space mission or the DECIGO space laser observatory).

Overall, this diagram shows that the dynamical equation \eqref{GW5} produces stable and physical responses for the above field parameters and clearly reveals the spectral dependence of gravitational waves on wave number.


\section{Conclusion}\label{VII}

In this study, an interactive HDE model was investigated in the modified gravity framework \(f(Q)\). By choosing a power law ansatz for the Hubble parameter \eqref{Hnt3} and using observational data \(H(z)\), the free parameters of the model were constrained and then the cosmological behavior of the model at the background level as well as in the tensor perturbations and gravitational waves section was studied. The statistical fitting results showed that the present model is able to reproduce the Hubble observational data with reasonable accuracy and provide values of \(H_0\simeq71.02~{\rm km~s^{-1}~Mpc^{-1}}\) and \(n\simeq1.015\) for its free parameters. Based on these values, the age of the universe is \(t_0=13.96~{\rm Gyr}\) which is in very good agreement with the new observational estimates and shows that the present model is able to describe the time scale of the evolution of the universe correctly.

Next, by introducing the effective critical density \(\rho_{\rm cr}^{(mod)}=\frac{1}{\kappa^2}\left(6f_QH^2-\frac{f}{2}\right)\) the relative density parameters of matter and dark energy were calculated. The results showed that the current values \(\Omega_{m0}=0.314\) and \(\Omega_{de0}=0.686\) are consistent with the accepted observational values with acceptable accuracy. Also, this definition in the limit \(f(Q)=Q\) exactly corresponds to the standard critical density of general relativity, which indicates the theoretical consistency of the model structure.

The study of the behavior of the energy density and pressure of dark energy showed that the density of dark energy remains positive throughout the history of the universe, while its pressure always has a negative sign. This behavior confirms that dark energy in the present model can play the role of a driving factor for the accelerated expansion of the universe. In addition, the parameter of the dark energy EoS at the present time has a value very close to \(\omega_{de}\simeq-1\) and in the past it tends to values greater than \(-1\). Therefore, dark energy in this model exhibits a behavior close to the cosmological constant at the present time and a quasi-quintessence behavior in the transition to higher redshifts.

On the other hand, the analysis of the effective speed of sound of dark energy showed that the model experiences a region of classical stability in the present era, although negative values for \(v_s^2\) appear in the transition to higher redshifts. This behavior is similar to the results reported in many HDE models and can be considered as an intrinsic feature of this class of models. Therefore, the negation of \(v_s^2\) is not necessarily a sign of model failure and should be evaluated along with other cosmological indicators.

Also, the deceleration parameter was obtained from the chosen ansatz structure directly as \(q=-1+\frac{1}{n}\) and for the best-fit values it has a negative value, indicating the current accelerated phase of the universe. However, the constancy of this quantity should be considered as a limitation resulting from the choice of the power law form of the Hubble parameter in the model reconstruction process and not as a fundamental property of the \(f(Q)\) theory.

An important part of this research was devoted to the study of gravitational waves. The results of solving the tensor perturbation equation showed that the amplitude of the gravitational wave exhibits a regular and oscillatory behavior during the evolution of the universe, and the presence of corrections due to \(f(Q)\) gravity and HDE has a noticeable effect on the propagation characteristics of gravitational waves. This indicates that gravitational waves can be used as an independent and powerful tool for testing HDE models in modified gravities.

Overall, the results of this study show that the interactive HDE model in the gravity framework \(f(Q)\) is able to provide a coherent picture of the evolution of the universe; in a way that not only does it successfully reproduce the main features of the universe at the background level, including the expansion history, the relative contribution of cosmic components, and the age of the universe, but also shows acceptable physical behavior in the tensor perturbation section. These results indicate that \(f(Q)\) gravity can provide a suitable platform for describing the nature of dark energy and investigating its effects on the propagation of gravitational waves, and therefore can be proposed as a promising framework for the study of modern cosmology and future observational experiments.

As a perspective for future research, a more detailed examination of cosmological perturbations, growth rates of structures, stability analysis at higher levels, and direct comparison of model predictions with gravitational wave observational data can impose more precise constraints on model parameters and provide a deeper understanding of the role of HDE in \(f(Q)\) gravity.

\bibliographystyle{apsrev4-2}
\bibliography{references}

\end{document}